\documentclass{ws-mpla}
\usepackage[super]{cite}
\usepackage{graphicx}
\begin{document}

\markboth{Bhattacharya, Thirukkanesh and Sharma}{Properties of relativistic star in $5$-D Einstein-Gauss-Bonnet gravity}

\catchline{}{}{}{}{}

\title{Properties of relativistic star in $5$-D Einstein-Gauss-Bonnet gravity}

\author{Soumik Bhattacharya}

\address{Department of Physics, Cooch Behar Panchanan Barma University, Cooch Behar 736101, India.\\
Email: soumik.astrophysics@gmail.com}

\author{Suntharalingam Thirukkanesh}

\address{Department of Mathematics, Eastern University, Chenkalady, Sri Lanka.\\
Email: thirukkanesh@esn.ac.lk
}

\author{Ranjan Sharma\footnote{Corresponding author}}

\address{Department of Physics, Cooch Behar Panchanan Barma University, Cooch Behar 736101, India.\\
Email: rsharma@associates.iucaa.in}

\maketitle

\pub{Received (Day Month Year)}{Revised (Day Month Year)}

\begin{abstract}
In recent years, there has been a growing interest in stellar modelling in the framework of Einstein-Gauss-Bonnet gravity. In this paper, for a relativistic star in static equilibrium, we invoke the $5$ dimensional Einstein-Gauss-Bonnet gravity and solve the system by assuming a matter distribution that admits a linear equation of state (EOS). We fix the model parameters by matching the interior solution to the exterior Boulware-Deser metric, which facilitates physical analysis of the resultant configuration. We analyze the star's gross physical properties, which brings to attention the role of the Gauss-Bonnet coupling parameter $\alpha$ in fine-tuning the values of the matter variables.

\end{abstract}

\ccode{PACS Nos.: }

\section{Introduction}
\label{intro}
 
The General Theory of Relativity (GTR) developed by Einstein relating spacetime curvature with matter is believed to be the correct theory of gravity till date. Despite the enormous success of Einstein's gravity (GR) in predicting several observational tests of gravity, including the most recent direct detection of gravitational waves from the merger of binary black holes and neutron stars by advanced LIGO and Advanced Virgo detectors to the direct imaging of black holes at the centres of Milky Way and $M87$ galaxies, a correct theory of gravity applicable at all scales has remained elusive till date. Among other issues such as the quantization of gravity, the theory comes short in explaining many observational phenomena, e.g., the late time accelerated expansion of the universe. In pursuit of alternative theories, GR is being modified in many ways, including modifications invoked by considering higher-order curvature terms in the Lagrangian. In the past, Kaluza and Klein \cite {Kalu21,Klein26} investigated the higher dimensional effects on electromagnetism (EM) by conjoining EM to gravity. A natural extension of GTR is the Lovelock gravity \cite {Love71}, which incorporates higher order curvature corrections to the field equations in a general higher dimensional spacetime. One obtains an effective theory of gravity for a quadratic form of the Lagrangian in Lovelock gravity, namely, the Einstein-Gauss-Bonnet (EGB) gravity. 

The Einstein-Gauss-Bonnet (EGB) gravity has got widespread attention primarily in cosmological models (see e.g., Ref.~ \refcite{Tan63,Chatter90,Gallo04,Ghosh08} and references therein). By modifying the Einstein- Hilbert action with the inclusion of a $4$D-Gauss-Bonnet term, Boulware and Deser \cite {Boul85} showed that black hole solutions might exist in $n \geq 5$ dimensions. In the recent past, in addition to its cosmological applications, there has been a growing interest in studying its astrophysical implications. It should be stressed that even though we live in a $4$-dimensional universe, such probes, particularly in the context of astrophysical systems, stem out of theoretical curiosity to understand the effects of higher order curvature corrections in a general $n \geq 4$ dimensions.

Our investigation is motivated by some recent developments in this field. Within the framework of $5$-D EGB gravity, Maharaj {\em et al} \cite{Maharaj15} developed an interior solution by assuming a barotropic perfect fluid distribution. Dadhich {\em et al} \cite{Dadhich10} established that the gravitational field inside a constant density fluid sphere has a universal character for the spacetime dimensions $n \geq 4$, which is not only valid for the Einstein-Hilbert action but also true for the more general Lovelock action. Recently, Tangphati {\em et al} \cite{Tang21} have developed a new class of solutions for a static spherically symmetric perfect fluid distribution composed of quark matter in EGB gravity. In a comparative study between EGB gravity and GTR \cite {Bhar17}, it has been claimed that some features of a stellar body such as energy conditions and causality behaviour might remain unaltered in EGB gravity. Effects of the EGB coupling term on stellar properties have been analyzed by developing various stellar models in the presence of the scalar field, anisotropy, bulk viscosity, shear, charge, and heat-flux \cite{Hans17,Bras20,Jasim21,Tang21}. Bhar {\em et al} \cite {Bhar19} developed a model of a static compact charged anisotropic fluid sphere possessing a linear relationship between the radial pressure and the energy density within the framework of Einstein-Maxwell-Gauss-Bonnet (EMGB) gravity. It has been shown that the mass and radius of a compact star for a given central density increase as the Coulomb repulsive force increases with the incorporation of electric charge \cite{Pret21}. It is noteworthy that Gauss-Bonnet gravity has also been studied in $n \rightarrow 4$ dimension. Some of the investigations include the incorporation of scalar fields such as dilations \cite{Tori97,Noji05,Chen10}, and rescaling of the coupling parameter \cite{Hans20}. The GB parameter has also been regularized in $4$ dimensions to generate a non-trivial gravitational theory \cite{Feng21}. It is, therefore, worthwhile to analyze the implications of EGB gravity on the geometry and physical behaviour of localized objects.

In our work, we study a star-like configuration by generating a new class of exact solutions within the framework of $5-$D EGB gravity. The stellar object is assumed to be filled with anisotropic matter that admits a linear equation of state (EOS). In the high-density regime, theoretical studies affirm that the radial and the tangential pressures need not be equal  \cite{Herr97}. For the assumed matter distribution, we make the EGB field equations tractable by introducing a particular coordinate transformation and closing the system in terms of a single generating function. We solve the system by specifying the generating function $Z(x)$. We show that the solution is well-behaved, stable and the energy conditions are satisfied. We fix the model parameters of the solution by matching the interior solution to the exterior Boulware-Deser metric \cite{Boul85}, which facilitates its physical analysis. Numerical analysis of the developed model helps us understand the effects of EGB gravity on the geometry and gross physical features of the resultant object.  
 
The paper is organized as follows: In section \ref{sec2}, the mathematical framework of EGB gravity is laid down. In section \ref{sec3}, we introduce a particular transformation to make the relevant field equations integrable. We solve the field equations for a linear EOS and express the physical quantities in analytic forms in section \ref{sec4}. In section \ref{sec5}, the interior solution is matched to the exterior Boulware-Deser solution. In section \ref{sec6}, we compare EGB and GR by analyzing the effects of the EGB coupling coefficient on the physical quantities. Some concluding remarks are made in section \ref{sec7}.

\section{EGB gravity: Basic framework}
\label{sec2}
In five dimensions, by incorporating the Gauss-Bonnet term, the Einstein-Hilbert action  can be written as
\begin{equation}\label{eq1}
S = \int \sqrt{-g} \left[\frac{1}{2}(R-2 \Lambda +\alpha L_{GB})\right]d^5x +S_{matter},
\end{equation}
where $\alpha$ is the Gauss-Bonnet coupling term originating from string theory. The corresponding equation of motion for the above action turns out to be a second-order quasi-linear equation. The strength of the action $L_{GB}$ can be understood for the Gauss-Bonnet term in 
dimensions $n\geq 5$. 
 
For the assumed action, the EGB field equations\footnote{in this 
formalism, we use geometric units with
 the coupling constant $\kappa$ set to unity}  are obtained as \cite {Maharaj15}:
\begin{equation}\label{eq2}
G_{ab}+\alpha H_{ab} = T_{ab},
\end{equation}
where $G_{ab}$ is the Einstein tensor. Note that we have considered the metric signature as (- + + + +). The Lanczos tensor has the form
\begin{equation}
H_{ab} =  2 (R R_{ab}-2R_{ac}R^c_b - 2R^{cd}R_{acbd}+R^{cde}_aR_{bcde})- \frac{1}{2}g_{ab}L_{GB}, \label{eq3}
\end{equation}
where the Lovelock term is given by 
\begin{equation}\label{eq4}
L_{GB} = R^2 +R_{abcd}R^{abcd}- 4R_{cd}R^{cd}.
\end{equation}

\section{Field equations}
\label{sec3}
 We couch the line element for the interior of a static anisotropic distribution of matter in $5$-dimensions (in coordinates $x^i = t,r,\theta,\phi,\psi$) in the standard form as 
 \begin{equation}
 ds^{2} =  -e^{2\nu(r)} dt^{2} + e^{2\lambda(r)} dr^{2} + r^{2} ( d\theta^{2} + \sin^{2}{\theta} d\phi^{2}+ \sin^{2}{\theta} \sin^{2}{\phi} d\psi^{2}), \label{eq5}
 \end{equation}
 where $\nu(r)$ and $\lambda(r)$ are the unknown gravitational potentials. The comoving fluid velocity is given by $u^i=e^{-\nu}\delta_0^i$
 and the unit space-like vector along the radial direction has the form $\theta ^i = e^{-\lambda}\delta_1^i$.
   
 The energy-momentum tensor of the anisotropic fluid is assumed to be of the form
 \begin {equation}
 \label{eq6} T_{ab}  =  (\rho + p_t)u_a u_b + p_t g_{ab} + (p_r - p_t) \theta_a \theta_b,
 \end {equation}
 where $\rho$ , $p_r$ and $p_t$ represent the matter density,
 radial and transverse pressure of the fluid, respectively.
 
 Subsequently, the field equations (\ref{eq2}) are obtained as
 \footnote{Here $(')$ denotes the differentiation with respect to the radial
 coordinate r}
 \begin{eqnarray}
 \label{eq6a}\rho &=& \frac{3}{e^{4\lambda }r^3} \left(4\alpha
 \lambda' +re^{2\lambda}-re^{4\lambda}- r^2 e^{2\lambda}\lambda' -4\alpha e^{2\lambda}\lambda'\right),\\
 \label{eq6b} p_r &=&\frac{3}{e^{4\lambda }r^3}
 \left[-re^{4\lambda}+
 (r^2 \nu' +r +4\alpha \nu')e^{2\lambda} -4\alpha \nu'\right], \\
 \label{eq6c} p_t &=& \frac{1}{e^{4\lambda }r^2} \left(- e^{4\lambda
 }- 4\alpha \nu''+ 12 \alpha \nu' \lambda' -4 \alpha
 (\nu')^2\right)
 \nonumber\\
 &+& \frac{1}{e^{2\lambda }r^2} \left(1- r^2 \nu' \lambda' +2r \nu'
 -2r  \lambda' +r^2(\nu')^2 \right) \nonumber \\
 &+&\frac{1}{e^{2\lambda }r^2} \left(r^2 \nu'' -4\alpha
 \nu'\lambda' + 4\alpha (\nu')^2 +4\alpha \nu''\right).
 \end{eqnarray}
 To  make the set of equations (\ref{eq6a})-(\ref{eq6c}) tractable, we make the following transformation 
 \begin{equation}
 \label{eq7} x =  r^2,~~ Z(x)  = e^{-2\lambda(r)} ~\mbox{and}~
 y^{2}(x) = e^{2\nu(r)}.
 \end{equation}
A similar transformation may be found in Ref.~\refcite{Dur83} and \refcite{Hans06}. Consequently, the system of equations (\ref{eq6a})-(\ref{eq6c}) takes the form:
  \begin{eqnarray}
 \label{eq8a}   \rho &=& 3\dot{Z} + \frac{3(Z-1)(1-4 \alpha \dot{Z})}{x}, \\
 \label{eq8b} p_r &=& \frac{3(Z-1)}{x}+\frac{\dot{y}}{y}
 \left[6Z- \frac{24\alpha (Z-1)Z}{x}\right]
 ,\\
 \label{eq8c} p_t &=& 2\left[3Z +x
 \dot{Z}-4 \alpha 
 \left(\frac{Z}{x}(Z-1)+(3Z-1)\dot{Z}\right)\right] \frac{\dot{y}}{y} \nonumber\\
 &+& 4Z \left[x-4\alpha 
 (Z-1)\right]\frac{\ddot{y}}{y} + \frac{Z-1}{x}+2 \dot{Z},
 \end{eqnarray}
 where a dot ($.$) denotes differentiation with respect to the variable $x$.
 
At this stage, we need to provide an equation of state (EOS) to close the system. Accordingly, we assume that the matter distribution satisfies the following linear barotropic EOS 
\begin{equation}
p_r = k\rho -B, \label{eq9}
\end{equation}
 where $k$ and $B$ are constants. It is to be noted that the MIT bag model EOS for ultra-compact stars composed of strange matter has a linear form. Zdunik \cite{Zdu00} has shown that a strange matter EOS can be approximated to a linearized form that exhibits a particular scaling behaviour of the system. Earlier, Sharma and Maharaj \cite {Sharma07} had assumed a linear EOS to develop a relativistic stellar model. All these provide us with enough motivation to choose a linear EOS to construct our model. Another advantage of choice (\ref{eq9}) is that it ensures that the radial pressure vanishes when the density reaches its surface value $\rho_s = B/k$. 
 
 Substituting (\ref{eq8a}) and (\ref{eq8b}) into (\ref{eq9}), we obtain
 \begin{equation}
 \frac{\dot{y}}{y} = \frac{3\left[k\dot{Z}x+(Z-1)[k(1-4\alpha \dot{Z})-1]\right]-B  x}{3[2xZ-8\alpha (Z-1)Z]}. \label{eq10}
 \end{equation}
 Integration of equation (\ref{eq10}) yields
 \begin{equation}
 y = d \exp \left[\int f(x)dx \right], \label{eq11}
 \end{equation}
 where $\displaystyle f(x)=\frac{3\left[k\dot{Z}x+(Z-1)[k(1-4\alpha
 \dot{Z})-1]\right]-B x}{3[2xZ-8\alpha (Z-1)Z]},$\\
  and $d$ is a
 constant of integration. Subsequently, we express the system of equations
 (\ref{eq8a})-(\ref{eq8c}) in the following form:  
 \begin{eqnarray}
 \label{eq12a}   \rho &=& 3\dot{Z} + \frac{3(Z-1)(1-4 \alpha  \dot{Z})}{x}, \\
 \label{eq12b} p_r &=& k\rho -B, \\
 \label{eq12c} p_t &=& \frac{Z-1}{x}+2 \dot{Z}+2 \left[3Z +x
 \dot{Z}-4 \alpha\left(\frac{Z}{x}(Z-1)+(3Z-1)\dot{Z} \right) \right] f(x) + \nonumber\\
 && 4Z \left[x-4\alpha (Z-1)\right]\left[\dot{f}(x)+(f(x))^2\right]
 \end{eqnarray}
 It is remarkable that the matter variables $\rho, p_r$ and
 $p_t$ are expressed in terms of the gravitational potential $Z$ only. We must choose $Z$ in such a manner that (i) equation (\ref{eq11}) becomes integrable and (ii) it provides a physically acceptable model. 
 
\section{Choosing $Z$} 
\label{sec4}
 Utilizing the additional degrees of freedom, we assume the gravitational potential $Z$ in the form
 \begin{equation}\label{eq13}
 Z = \frac{(1+2ax)}{(1+ax)},
 \end{equation}
 where $a$ is a real constant. The objective of the specific choice is to ensure that the system becomes integrable, and to the best of our knowledge, this particular choice has not been used earlier to develop a stellar model in EGB gravity. 
With this assumption, the physical quantities take the following forms:
 \begin{eqnarray}
 \label{eq14a} e^{2\nu} &=& d^2 (1+2ax)^l (1+ax)^{-k}[1+ax-4a\alpha]^h\exp{\left[- \frac{B(1+ax)}{6a}\right]},\\
 \label{eq14b} e^{2\lambda} &=& \frac{(1+ax)}{(1+2ax)}, \\
 \label{eq14c} \rho &=&  \frac{3 a  (2 + 3 a x + a^2 x^2 - 4 a \alpha)}{(1+ax)^3}, \\
 \label{eq14d} p_r &=& \frac{3 a  k (2 + 3 a x + a^2 x^2 - 4 a 
 \alpha)}{(1+ax)^3}-B,\\
 \label{eq14e} p_t &=& \frac{a(3 + a x)}{(1 + a x)^2}+
 \frac{2}{(1 + a x)^3}\left[3 + a (13 x - 12 \alpha) \right.\nonumber\\ 
 &&\left. + 2 a^3 x^2 (3 x - 4 \alpha) + 
16 a^2 x (x - 2 \alpha)\right] f(x) +\nonumber\\
 && \left[\frac{4 (1 + 2 a x) [x + a x (x - 4\alpha)]}{(1 + a x)^2} \right] \left[\dot{f(x)}+ f(x)^2\right],
 \end{eqnarray}
 where
 \begin{eqnarray}
 l&=& \frac{B + 6a[1+k(16a \alpha -3)]}{12 a (8a
 \alpha  -1)}, \nonumber\\
 h &= & \frac{4a \alpha [3(1 - k) + 4 B \alpha)}{3(1-8 a
 \alpha)},
 \nonumber\\
 f(x) &=&- \left[B  + a^3 x^2 (3 - 3 k + B  x) + a (3 - 6 k + 3 B x^2) +3 a^2 ((2 - 3 k) x + Bx + 4  k \alpha)\right]/
 \nonumber\\
 && 6 (1 + a x) (1 + 2 a x) [1 + a (x - 4  \alpha)]. \nonumber
 \end{eqnarray} 
 
\section{Exterior spacetime and matching conditions} \label{sec5}
The exterior spacetime corresponding to the $5$-dimensional static distribution is assumed to be described by the Boulware-Deser \cite{Boul85} metric 
 \begin{equation}
 ds^{2} = -F(r) dt^{2} + [F(r)]^{-1} dr^{2} +  r^2(d\theta^{2} + \sin^{2}{\theta} d\phi^{2}+ \sin^{2}{\theta} \sin^{2}{\phi} d\psi^{2}),\label{eq15}
 \end{equation}
with 
 \begin{equation}
 F(r) = \eta +\frac{r^2}{4 \alpha}
 \left(1-\sqrt{1+\frac{8\alpha M}{r^4}}\right), \label{eq16}
 \end{equation}
where  $M$ is the gravitational mass and $\eta$ is an arbitrary constant. Note that, with $\eta = 0,~\pm 1,$ the line element (\ref{eq15}) corresponds to a cosmological brane world solution \cite{Kraus99,Ida00,Dav03} in EGB gravity. Moreover, with $\eta = 1$,  the solution asymptotically corresponds to the Schwarzschild solution with a positive gravitational mass or Schwarzschild-de Sitter solution with a negative gravitational mass \cite{Abb82}. 
    
To fix the values of the model parameters, we first match the gravitational potentials of the interior metric with the exterior metric at the boundary $r = R$. Matching of $g_{rr}$ determines $a$ in the form
\begin{equation}
a = \frac{(1 - \eta) - A}{ R^2 [A + (\eta - 2)]}, \label {eq17}
\end{equation}
where
\begin{equation}
A = \frac {R^2}{4 \alpha} [1 - \sqrt{1+\frac{8\alpha M}{R^4}}]. \label {eq18}
\end{equation}

Utilizing the requirement that radial pressure must vanish at the boundary, we obtain $B$ as
\begin{equation}
B = \frac{3 a k (2 + 3 a R^2 + a^2 R^4 - 4 a \alpha)}{(1 + a R^2)^3}. \label{eq18b}
\end{equation}
 
Matching of $g_{tt}$ across the boundary yields
 \begin{equation}
 d = \pm   \left[\frac{F(R)(1 + aR^2)^k}{(1 + 2aR^2)^l(1 + aR^2 - 4a\alpha)^h}\right]^{1/2}\exp{\left[- \frac{B(1+aR^2)}{12a}\right]}, \label{eq19}
 \end{equation}
 where 
 \begin{equation}
 F(r=R) = \eta + A. \label{eq19a}   
 \end{equation}
It is interesting to note that equations (\ref{eq17}) and (\ref{eq19}) determine the constants $a$ and $d$ in terms of the radius of the star $R$ and gravitational mass $M$. To get an estimate of the values of the model parameters, for a given mass and radius, values of different physical quantities are given in Table~\ref{tab1} where following Bhar and Govender \cite{Bhar19}, we set $\eta = 1.5$ and $k=0.35$ in our calculation. 

\begin{table}[h] 
\tbl{Values of physical parameters for different $\alpha$ ($R=10 $, $k=0.35$, $M=2$).}
{\begin{tabular}{@{}cccc@{}} \toprule
$\alpha$ & $a$ & $\rho_c$ & $B$  \\
\colrule
$0.01$    &  $0.00923077$     &   $0.0553744$      &  $0.0076603$ \\
$0.1$     &  $0.0092308$      &   $0.0552825$      &  $0.00765578$ \\
$1.0$     &  $0.00923106$     &   $0.0543638$      &  $0.00761057$ \\
$5.0$     &  $0.00923224$     &   $0.0502794$      &  $0.00740962$ \\
$10.0$    &  $0.0092337$      &   $0.0451709$      &  $0.00715838$ \\
$15.0$    &  $0.00923516$     &   $0.0400591$      &  $0.0069071$ \\
$20.0$    &  $0.0092366$      &   $0.034944$       &  $0.00665578$ \\
$25.0$    &  $0.00923802$     &   $0.0298258$      &  $0.0064044$ \\
\hline
\end{tabular}\label{tab1}}
\end{table}

\subsection{Choice of $\alpha$}
In equation (\ref{eq14c}), by setting $x=0$, we obtain the central density  
\begin{equation}
\rho_{c} = 3a(2-4a \alpha). \label{eq19b}
\end{equation}
Now, the central density will remain positive if either of the following two conditions is satisfied:\\
i) $a > 0$ \& $(2-4a \alpha)>0$, or ii) $a<0$ \& $(2-4a \alpha)<0$. \\
Condition (i) implies $\alpha < \frac{1}{2a}$ and condition (ii) implies $\alpha > \frac{1}{2a}$. 

Therefore, in our model, the signature of  $\alpha$ depends on the choice of the curvature parameter $a$ appearing in the metric ansatz (\ref{eq13}). For numerical analysis, we consider only positive values of $a$. While a negative value of the coupling parameter is permissible (see e.g.,  Ref.~ \cite{Dey22}), it is, in general, restricted to positive coupling only as can be seen in references \cite{Gross87,Bento96,Maeda06}. A positive $\alpha$ is consistent with its physical meaning as the inverse string tension having a dimension of $[$Length$]^2$  in string theory \cite{Tang21}.

\section{Physical features} \label{sec6}

For a fixed set of model parameters, we now utilize the solution to analyze the physical features of the stellar configuration. Some salient features of the resultant configuration are discussed below:
 
\begin{enumerate}
\item We plot the matter variables in Fig.~(\ref {f1}), (\ref {f2}) and (\ref {f3}). The plots show that the matter density $\rho$, the radial pressure $P_{r}$ and the tangential pressure $P_{t}$ are all positive and decrease from their peak values at the centre towards the surface as expected for a physically acceptable model. Interestingly, for a given mass and radius, the density and radial pressure consistently decrease for higher Gauss-Bonnet coupling ($\alpha$) values. The tangential pressure towards the centre also exhibits similar behaviour.
 
\begin{figure}[h]
\centering
\begin{minipage}{0.49\textwidth}
\centering
\includegraphics[width=1\textwidth]{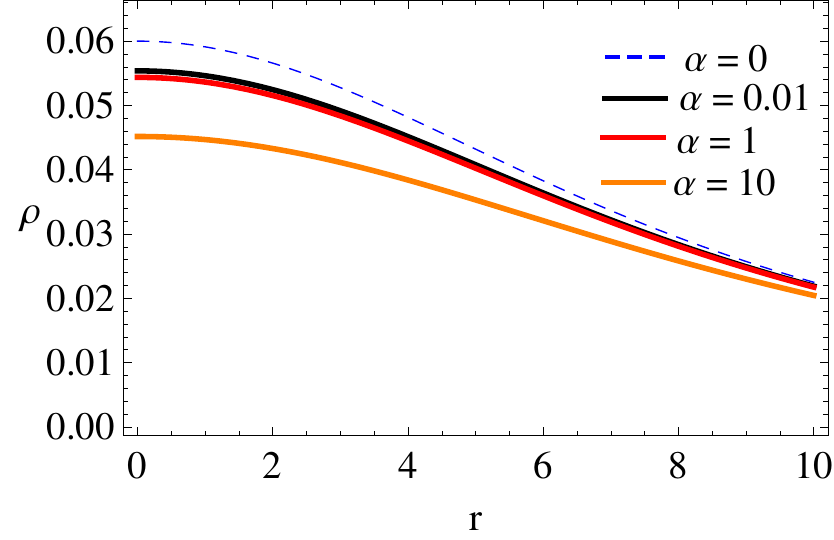}
\caption{Matter density $\rho$ plotted against the radial distance $r$. (We have assumed $R = 10,~ M = 2,~ \eta = 1.5,~ k = 0.35$)}
\label{f1} 
\end{minipage}
\hfill
\begin{minipage}{0.49\textwidth}
\centering
\includegraphics[width=1\textwidth]{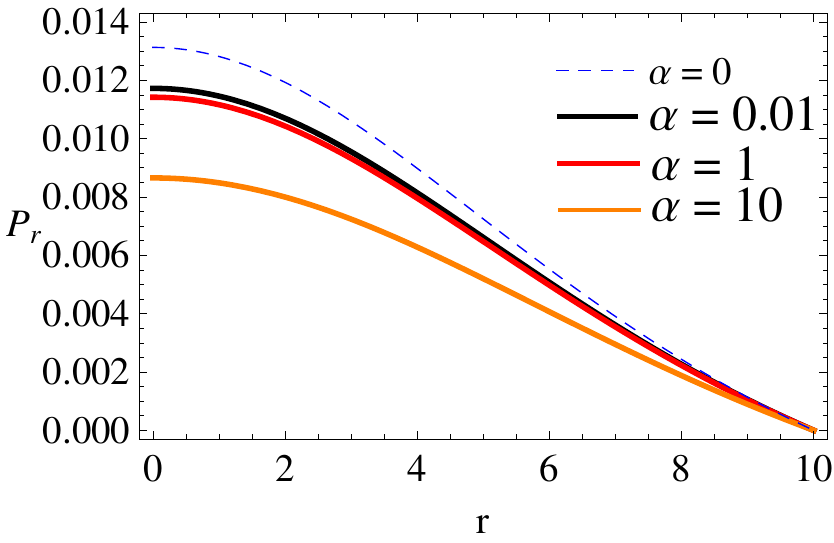}
\caption{Radial pressure $P_{r}$ plotted against the radial distance $r$ by taking the same values of the parameters mentioned in Fig.~ \ref {f1}.}
\label{f2}   
\end{minipage}
\end{figure}

\item 
The radial variation of anisotropy ($\displaystyle \Delta = P_{t} - P_{r} $) is shown in Fig.~\ref{f4}. We  note that both $P_{t}$ and $P_{r}$ are equal at the centre, as expected. For $\alpha \neq 0$, $\Delta$ reaches a maximum value and then decreases. 

\begin{figure}[h]
\centering

\begin{minipage}{0.49\textwidth}
\centering
\includegraphics[width=1\textwidth]{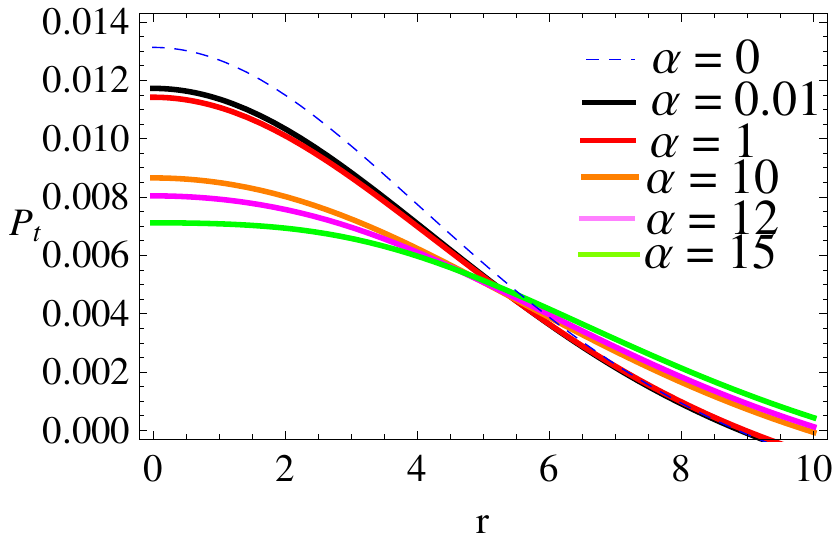}
\caption{Tangential pressure $P_{t}$ plotted against the radial distance $r$ by taking the same values of the parameters mentioned in Fig.~ \ref {f1}.}
\label{f3} 
\end{minipage}
\hfill
\begin{minipage}{0.49\textwidth}
\centering
\includegraphics[width=1\textwidth]{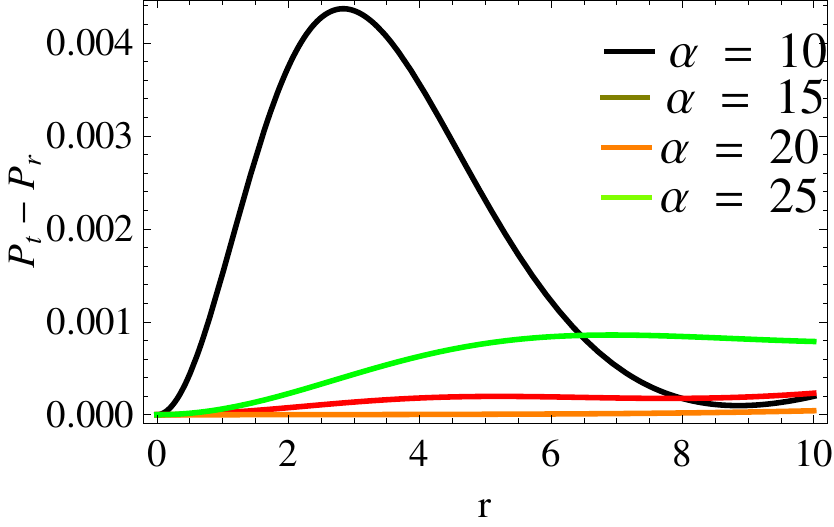}
\caption{Anisotropy $P_{t} - P_{r}$ plotted against the radial distance $r$ by taking the model parameters as: $M = 2$, $R = 10$, $k = 0.35$, $(i)$ $\eta = 1.7$ for $\alpha = 10$, $(ii)$ $\eta = 1.43$ for $\alpha = 15$, $(iii)$ $\eta = 1.22$ for $\alpha = 20$, $(iv)$ $\eta = 1.4$ for $\alpha = 25$}
\label{f4}       
\end{minipage}
\end{figure}

 \item Fulfillment of energy conditions:  A realistic star should fulfill the following energy conditions:
  
 \begin{itemize}
     \item Null Energy Condition (NEC): $\rho \geq 0$.
 \end{itemize}
 \begin{itemize}
     \item Weak Energy Condition (WEC) : $\rho - P_{r} \geq 0 , ~ \rho - P_{t} \geq 0$.
 \end{itemize}    
 \begin{itemize}
     \item Strong Energy Condition (SEC): $\rho - P_{r} - 2P_{t} \geq 0 $.
 \end{itemize}   
 
Fig.~(\ref{f5})-(\ref{f7}) clearly show fulfillment of these conditions in this model.

\begin{figure}[h]
\centering
\begin{minipage}{0.49\textwidth}
\centering
\includegraphics[width=1\textwidth]{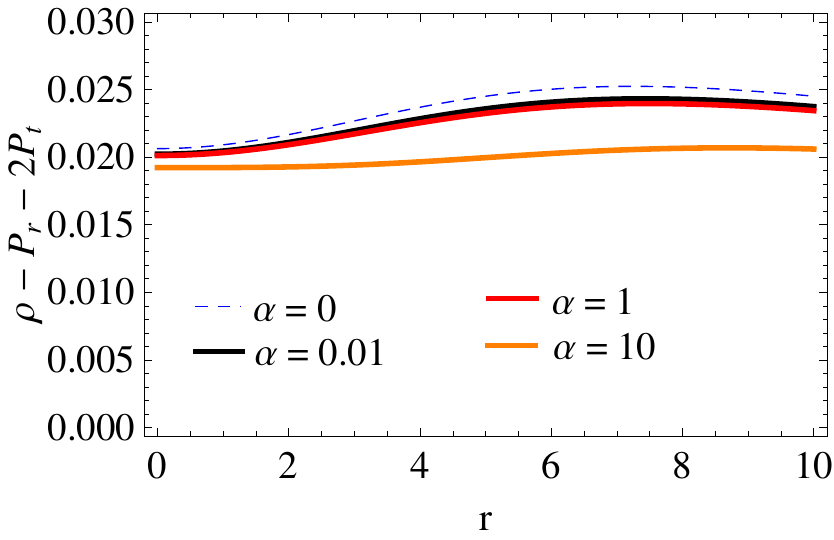}
\caption{$\rho - P_{r} - 2P_{t}$ plotted against the radial distance $r$ by taking the same values of the parameters mentioned in Fig.~\ref{f1}.}
\label{f5}
\end{minipage}
\hfill
\begin{minipage}{0.49\textwidth}
\centering
\includegraphics[width=1\textwidth]{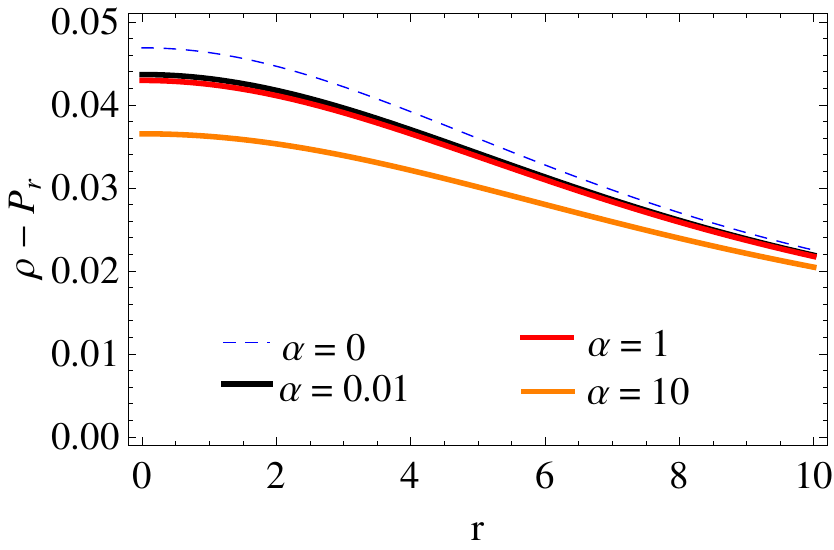}
\caption{$\rho - P_{r} $ plotted against the radial distance $r$ by taking the same values of the parameters mentioned in Fig.~\ref {f1}.}
\label{f6} 
\end{minipage}
\end{figure}

  \item
In this model, since we have assumed a linear relationship between the density and radial pressure given in equation (\ref{eq9}), the square of the radial sound velocity ($v_{sr}^2$) is simply a constant $k$ in the presence or absence of the coupling term. However, the square of transverse sound velocity ($v_{st}^2$) shows a variation that falls off with the radial distance $r$ and $\alpha$ as shown in Fig.~\ref{f8}.

\begin{figure}[h]
\centering
\begin{minipage}{0.49\textwidth}
\centering
\includegraphics[width=1\textwidth]{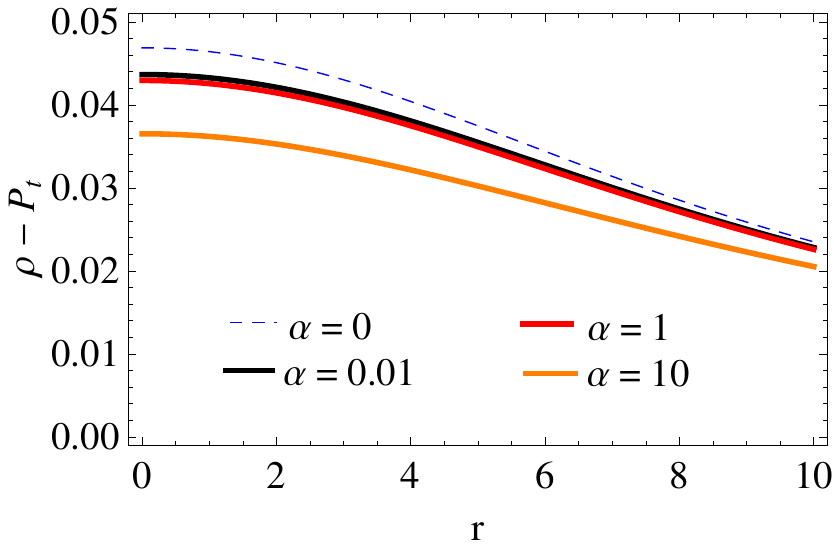}
\caption{$\rho - P_{t}$ plotted against the radial distance $r$ by taking the same values of the parameters mentioned in Fig.~\ref {f1}.}
\label{f7} 
\end{minipage}
\hfill
\begin{minipage}{0.49\textwidth}
\centering
\includegraphics[width=1\textwidth]{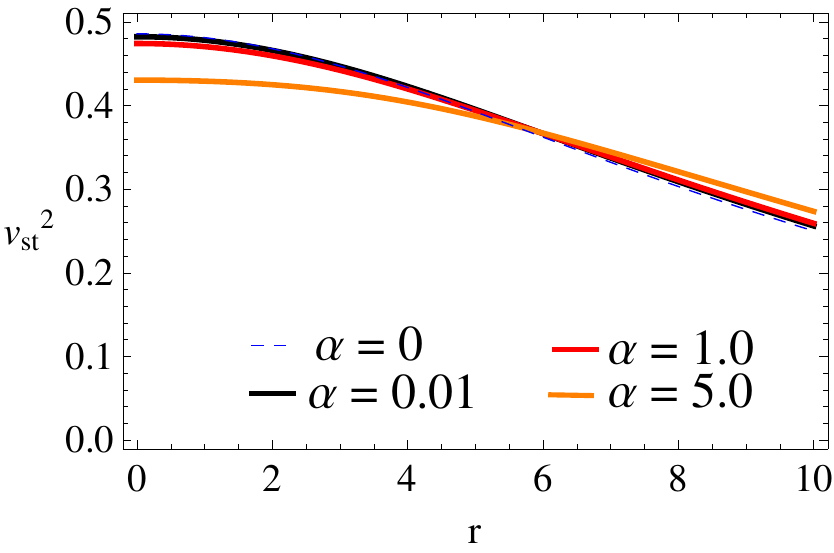}
\caption{Transverse sound velocity $v_{st}^2$ plotted against the radial distance $r$ by taking the same values of the parameters mentioned in Fig.~ \ref {f1}.}
\label{f8}
\end{minipage}
\end{figure}

   \item
  An important measure of a stellar object is the adiabatic index which is defined as 
 \begin{equation}
 \Gamma = \frac{\rho + P_{r}}{P_{r}} \frac{dP_{r}}{d\rho},\label{20}
 \end{equation}
For a stable stellar configuration, $\Gamma$ should be greater than $4/3$ everywhere within the stellar interior \cite{Hein75}. This condition is shown to be satisfied in this model in Fig.~\ref{f9}. A marginal increment in the values of $\Gamma$ is noted for higher values of $\alpha$.

\begin{figure}[h]
    \centering
\begin{minipage}{0.55\textwidth}
\centering
\includegraphics[width=1\textwidth]{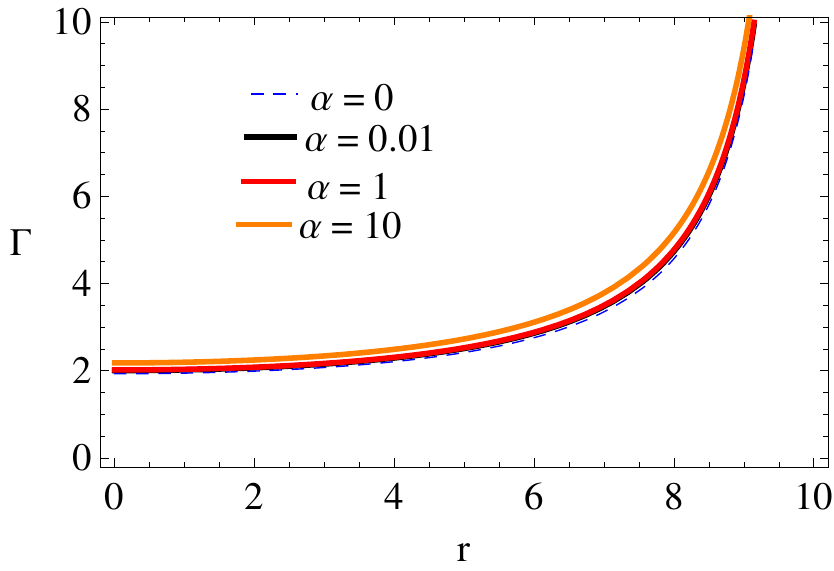}
\caption{Adiabatic index $\Gamma$ plotted against radial distance $r$ by taking the same values of the parameters mentioned in Fig.~ \ref {f1}.}
\label{f9} 
\end{minipage}
\end{figure}

 \item 
To examine stability of the configuration, we have applied the `cracking' condition of \cite{Herrera92}. In Fig.~\ref{f10}, we note that the difference between the radial sound speed ($v_{sr}^2$) and the tangential sound speed ($v_{st}^2$) becomes zero at a certain radial distance which shifts outwards with the increase of $\alpha$ without violating the overturning (cracking) condition $0 < \left|v_{st}^2 - v_{sr}^2\right|~ < 1$.  

\begin{figure}[h]
\centering
\begin{minipage}{0.55\textwidth}
\centering
\includegraphics[width=1\textwidth]{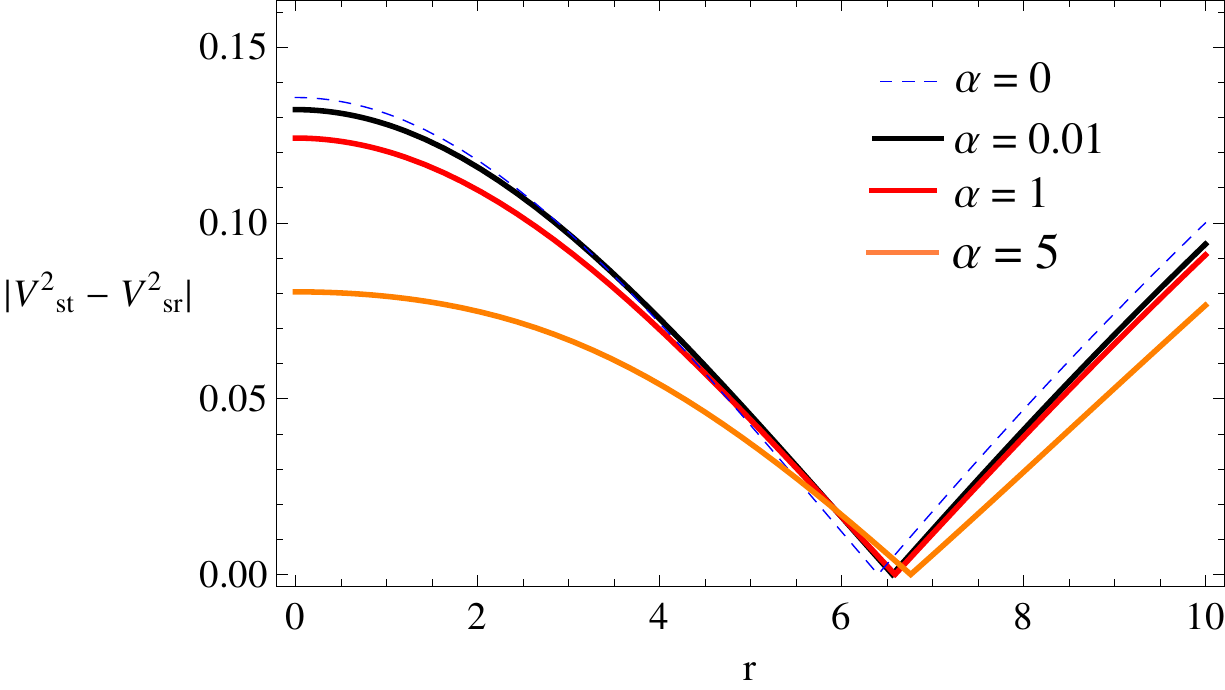}
\caption{Fulfillment of Harrera's cracking condition.}
\label{f10} 
\end{minipage}
\end{figure}

\end{enumerate}

\section{Conclusion}
\label{sec7}
We have generated a new solution describing a relativistic stellar configuration in Einstein-Gauss-Bonnet gravity which is regular, well-behaved and physically reasonable. The solution facilitates an analysis of the Gauss-Bonnet coupling term on the gross physical properties of the star. In particular, we have analyzed the role of the coupling term $\alpha$ on the physical quantities of the star. In this context, we would like to point out the following:
\begin{itemize}
\item The energy density and radial pressure decrease with the increase of the EGB coupling parameter $\alpha$. This feature is more pronounced at the central region.
\item While the tangential pressure decreases with the increase of the EGB coupling parameter $\alpha$ at the central region, an inverse relationship is observed towards the surface.
\item Energy conditions are satisfied for a wide range of values of $\alpha$. The stability of the configuration remains unaltered for a wide range of values of $\alpha$.
\end{itemize}

Our results points towards the fact the the Gauss-Bonnet gravity might play an important role in fine-tuning stellar observables. Further progress in this direction will be reported elsewhere.

\section*{Acknowledgments}
RS gratefully acknowledge support from the Inter-University Centre for Astronomy and Astrophysics (IUCAA),
Pune, India, under its Visiting Research Associateship Programme. We express our gratitude to the anonymous referee for his comments and suggestions.

\section*{Data availability} The manuscript has no associated data.

\end{document}